\documentstyle[emulateapj,psfig]{article}

\received{21 June 2000}
\accepted{29 August 2000}

\def\l{$\lambda$}
\def\degr{\hbox{$^\circ$}}
\def\arcmin{\hbox{$^\prime$}}
\def\arcsec{\hbox{$^{\prime\prime}$}}

\def\lsim{\mathrel{\hbox{\rlap{\lower.55ex \hbox {$\sim$}}\kern-.0em
\raise.4ex \hbox{$<$}}}} 
\def\gsim{\mathrel{\hbox{\rlap{\lower.55ex \hbox {$\sim$}}\kern-.0em
\raise.4ex \hbox{$>$}}}} 
\newcommand{\gpm}[3]{$#1^{+#2}_{-#3}$}

\slugcomment{Accepted for publication in The Astrophysical Journal}
 
\begin{document}

\title{VLT spectroscopy of GRB\,990510 and GRB\,990712; probing the
faint and bright end of the GRB host galaxy population \footnote{Based
on observations collected at the European Southern Observatory, Chile;
proposal no. 63.O-0567}}

\author{P.M. Vreeswijk\altaffilmark{1},
A. Fruchter\altaffilmark{2},
L. Kaper\altaffilmark{1},
E. Rol\altaffilmark{1},
T.J. Galama\altaffilmark{3},
J. van Paradijs\altaffilmark{1,4}\footnote{deceased},
C. Kouveliotou\altaffilmark{5,6},
R.A.M.J. Wijers\altaffilmark{7}, 
E. Pian\altaffilmark{8},
E. Palazzi\altaffilmark{8},
N. Masetti\altaffilmark{8},
F. Frontera\altaffilmark{8,9},
S. Savaglio\altaffilmark{2,10,11},
K. Reinsch\altaffilmark{12},
F.V. Hessman\altaffilmark{12},
K. Beuermann\altaffilmark{12},
H. Nicklas\altaffilmark{12},
E.P.J. van den Heuvel\altaffilmark{1}}	 

\altaffiltext{1}{Astronomical Institute `Anton Pannekoek', University
of Amsterdam \& Center for High Energy Astrophysics, Kruislaan 403,
1098 SJ Amsterdam, The Netherlands}

\altaffiltext{2}{Space Telescope Science Institute, 3700 San Martin Drive,
Baltimore, MD 21218, USA}

\altaffiltext{3}{California Institute of Technology, 1200 East
California Boulevard, Pasadena, California 91125}

\altaffiltext{4}{Physics Department, University of Alabama in
Huntsville, Huntsville AL 35899, USA}

\altaffiltext{5}{Universities Space Research Association}

\altaffiltext{6}{NASA/MSFC, Code SD-50, Huntsville AL 35812, USA}

\altaffiltext{7}{Department of Physics and Astronomy, SUNY Stony
Brook, NY 11794-3800, USA}

\altaffiltext{8}{ITESRE-CNR Bologna, Via P. Gobetti 101, 40129 Bologna, Italy}

\altaffiltext{9}{Physics Department, University of Ferrara, Via Paradiso,
12, 44100 Ferrara, Italy}

\altaffiltext{10}{On assignment from the Space Science Department of
the European Space Agency}

\altaffiltext{11}{Currently at the Observatory of Rome, via di
Frascati 33, I-00040 Monteporzio Catone, Italy}

\altaffiltext{12}{Universit\"ats-Sternwarte, Geismarlandstr. 11, D-37083
G\"ottingen, Germany}

\setcounter{footnote}{0}
\begin{abstract}

We present time-resolved optical spectroscopy of the afterglows of the
gamma-ray bursts GRB\,990510 and GRB\,990712. Through the
identification of several absorption lines in the first epoch
GRB\,990510 spectrum, we determine the redshift for this burst at $z
\geq$ 1.619. No clear emission lines are detected. The strength of the
Mg{\sc i} feature is indicative of a dense environment, most likely
the host galaxy of GRB\,990510. Although the host is extremely faint
(V $\gsim$ 28), the GRB afterglow allows us to probe its interstellar
medium and -- in principle -- to measure its metallicity.  The optical
spectrum of GRB\,990712 (whose host galaxy is the brightest of the
known GRB hosts at cosmological redshifts), shows clear features both
in emission and absorption, at a redshift of $z$ = 0.4331 $\pm$
0.0004. On the basis of several line emission diagnostic diagrams, we
conclude that the host galaxy of GRB\,990712 is most likely an H{\sc
ii} galaxy. We derive a unreddened [O{\sc ii}] star formation rate of
2.7 $\pm$ 0.8 M$_{\odot}$ yr$^{-1}$. Correcting for the measured
extinction intrinsic to the host galaxy (A$_V$ = \gpm{3.4}{2.4}{1.7}),
this value increases to \gpm{35}{178}{25} M$_{\odot}$ yr$^{-1}$. The
[O{\sc ii}] equivalent width, compared to that of field galaxies at $z
\leq$ 1, also suggests that the host of GRB\,990712 is vigorously
forming stars. We employ the oxygen and H$\beta$ emission-line
intensities to estimate the global oxygen abundance for the host of
GRB\,990712: log(O/H) = --3.7 $\pm$ 0.4, which is slightly below the
lowest metallicity one finds in nearby spiral galaxies. For both GRBs
we study the time evolution of the absorption lines, whose equivalent
width might be expected to change with time if the burst resides in a
dense compact medium. We find no evidence for a significant change in
the Mg{\sc ii} width.

\end{abstract} 


%\notetoeditor{} 

\section{Introduction}

In February 1997, the Italian-Dutch satellite {\it BeppoSAX} enabled a
breakthrough in the understanding of gamma-ray bursts (GRBs) by
providing an accurate position for the prompt X-ray emission of a
GRB. This led to the discovery of the first X-ray afterglow of a GRB
(Costa et al. 1997), and, independently, to the identification of the
first optical counterpart of a burster (Van Paradijs et al.
1997). Since then, several X-ray, optical and radio counterparts of
GRBs have been detected.  These multi-wavelength afterglow
observations can be explained reasonably well by simple fireball
models (for recent reviews see Piran 1999, and Van Paradijs,
Kouveliotou \& Wijers 2000). GRB distance determinations are crucial
in the effort to establish the physical nature of their
progenitor(s). The observed redshift distribution of the `normal'
afterglows (i.e., excluding GRB\,980425, which is associated with
supernova SN1998bw at z = 0.0085; Galama et al. 1998), ranges from $z$
= 0.43 (Galama et al. 1999, and this paper) to $z$ = 3.42 (Kulkarni et
al. 1998).

Although major advances in the understanding of GRBs have been made
over the past few years (thanks to the detection of afterglows), the
physical nature of their progenitor(s) remains unclear. The most
popular models are (i) the collapse of a rotating massive star
(Woosley 1993, MacFadyen \& Woosley 1999) and (ii) the merging of two
neutron stars, or a neutron star and a black hole (Narayan,
Paczy\'nski \& Piran 1992, Janka et al. 1999). The former
(`collapsar') model has trouble producing GRBs with durations shorter
than a couple of seconds and predicts that every afterglow is
accompanied by a supernova of a type (Ic) similar to SN1998bw
(MacFadyen \& Woosley 1999). Furthermore, in the collapsar environment
it is likely that the optical light of the afterglow is heavily
absorbed by the surrounding dusty medium. Given the massive
progenitors, one expects the frequency of GRBs to be strongly
correlated with the cosmic star formation rate; the latter remains one
of the great unresolved issues in astronomy of today.  The compact
star merger scenario can make short GRBs as well as long ones
(although a 10$^{15}$ G magnetic field is probably needed for the
latter, see M\'esz\'aros 2000); some of these mergers are expected to
occur in low-density environments, possibly located several kiloparsec
outside their host galaxies. This is due to the large kick velocities
imparted to the compact objects from the two respective supernovae
($\sim$ 250 -- 300 km s$^{-1}$; Hansen \& Phinney 1997), combined with
the long time between the birth of the system and the merger
occurrence (10$^8$ -- 10$^9$ years; Portegies Zwart \& Yungelson
1998).  Consequently, since the optical afterglow brightness depends
on the density of the circumsource medium, some of these bursts may
not show an afterglow at all. These could account for the `dark' burst
population, i.e. bursts for which only X-ray afterglows have been
found. One way to discriminate between these two models is by studying
the immediate environment of the burst. In the collapsar model the
circumsource density is expected to drop with distance as r$^{-2}$,
due to the expanding stellar wind of the SN progenitor, while in the
binary merger scenario a constant, relatively low-density ambient
medium is most plausible.

If the GRB source resides in a compact, gas-rich environment (which is
expected in the collapsar scenario), the afterglow spectrum might show
time-dependent absorption features (such as Ly$\alpha$ and Mg{\sc ii})
due to the gradual ionization of the surrounding medium (Perna \& Loeb
1998). In this case a decrease of the absorption-line equivalent
widths (EWs) with time is expected. On the other hand, spectroscopic
observations of the star HD 72089, situated behind the Vela supernova
remnant, show an increase of an order of magnitude of the absorption
strengths of elements such as Al and Fe, over the velocity range
spanned by absorption in the remnant (Jenkins \& Wallerstein 1995; see
also Savage \& Sembach 1996). This is attributed to the destruction of
the dust grains, due to the propagation of the SN shock, which causes
the release of elements (such as Fe and Mg) that are frozen in the
dust. Thus in a dusty environment that is being `shocked' by a GRB
explosion, one might expect the strength of the absorption lines to
increase in time.

In order to test these theories of GRB genesis, and the effects of
GRBs on their environments, we have an on-going program to obtain
spectra of GRB afterglows using the {\it Very Large Telescope} (VLT)
of the European Southern Observatory (ESO) at Paranal, Chile. Here we
present results on two of these bursts: GRB\,990510 and GRB\,990712.

\subsection{GRB\,990510}

GRB\,990510 was observed on 1999 May 10.36743 UT with the {\it Wide
Field Camera} (WFC) unit 2 onboard {\it BeppoSAX}, which localized the
burst at R.A. = 13$^h$38$^m$06$^s$, Decl. = --80\degr 29\arcmin
30\arcsec\ (J2000.0), with an error radius of 3\arcmin\ (Dadina et
al. 1999). The {\it BeppoSAX} {\it Gamma Ray
Burst Monitor} (GRBM) recorded an 80s event with a multi-peak
structure (40-700 keV). The average and peak intensity in the WFC unit
2 (2-28 keV) was about 0.7 and 4.3 Crab, respectively. The burst
position as determined with {\it BeppoSAX} is consistent with that of
the {\it Inter-Planetary Network} (IPN; Hurley et al. 1999), using the
{\it Burst And Transient Source Experiment} (BATSE) onboard the {\it
Compton Gamma Ray Observatory}, and {\it Ulysses}. BATSE recorded a
fluence above 20 keV of 2.56 $\times$ 10$^{-5}$ erg cm$^{-2}$ (Kippen
et al. 1999), ranking it in the top 9\% of the burst fluence
distribution.

We discovered the optical afterglow on images taken at the South
African Astronomical Observatory (SAAO) 1m telescope (Vreeswijk et
al. 1999) and subsequently triggered our VLT program to take spectra
and polarimetric images. Here, we present the time-resolved
spectroscopy. Our polarimetric observations resulted in the first
polarization detection of a GRB afterglow (Wijers et al. 1999; see
also Covino et al. 1999), while our photometric observations show an
achromatic break in the BVRIJHK light curves, which is most likely due
to the burst emission being collimated (Rol et al. 2000a; see also
Stanek et al. 1999, and Harrison et al. 1999). Fruchter et al. (1999)
have used HST to estimate V$_{\rm host}$ $\gsim$ 28, and do not find
evidence for a supernova (SN) of the same type and brightness as
SN\,1998bw in the late-time light curve of GRB\,990510. Recent HST
observations (April 2000) appear to detect a faint galaxy (V $\sim$
28) at the position of the early optical transient, which, if real, is
most likely the host of GRB\,990510 (Fruchter et al. 2000, Bloom et
al. 2000).

\begin{deluxetable}{ccccccc}
\footnotesize
\tablecaption{Log of the GRB\,990510 and GRB\,990712 spectroscopic
observations. \label{tab:log}}
\tablecomments{The total exposure time for all spectra was 1800 sec.}
\tablehead{
\colhead{UT date	(1999)} & 
\colhead{time since burst} &
\colhead{grism} &
\colhead{wavelength range} &
\colhead{resolving power} &
\colhead{S/N at 6500 \AA} \\
\colhead{} & 
\colhead{(days)} &
\colhead{} &
\colhead{(\AA)} &
\colhead{} &
\colhead{}
}
\startdata
\underline{GRB\,990510} & & & & & & \\
May 11.179 &   0.811  & G150I 			 & 3700-7700 & 185 & 91 \\
May 11.203 &   0.836  & G150I+OG590 		 & 6000-9000 & 280 & 78 \\
May 12.123 &   1.755  & G150I			 & 3700-7700 & 185 & 6 \\
May 12.146 &   1.779  & G150I+OG590		 & 6000-9000 & 280 & 6 \\
May 14.273 &   3.906  & G300V and G300I		 & 3880-9255 & 420 and 680 & 9\\
May 16.254 &   5.887  & G150I 			 & 3700-7700 & 185 & 5 \\
\underline{GRB\,990712} & & & & \\			     
July 13.182 & 0.485  & G150I			 & 3700-7700 & 185 & 26 \\
July 13.421 & 0.725  & G150I			 & 3700-7700 & 185 & 21 \\
July 14.181 & 1.485  & G150I			 & 3700-7700 & 185 & 14 \\
\enddata
\end{deluxetable}

\subsection{GRB\,990712}

GRB\,990712 triggered the GRBM and WFC unit 2 onboard {\it BeppoSAX}
on 1999 July 12.69655 UT. The burst lasted for about 30s, had a
double-peaked structure, was moderate in $\gamma$ rays, and was
accompanied by one of the strongest prompt X-ray counterparts ever
observed (Heise et al. 1999). The WFC unit 2 located the burst at
R.A. = 22$^h$31$^m$50$^s$, Decl. = --73\degr 24\arcmin 24\arcsec\
(J2000.0), with an error radius of 2\arcmin. Unfortunately, neither
flux nor fluence levels are reported in the literature. Again the SAAO
1m telescope was successful in hunting down the GRB afterglow (Bakos
et al. 1999), which allowed us to quickly alert the VLT staff for
spectroscopic, polarimetric and further photometric follow-up
observations. The host galaxy of GRB\,990712 is the brightest of the
known GRB host galaxies, with R = 21.8 and V = 22.3 (Sahu et
al. 2000).  The VLT polarimetric images exhibit a significant degree
of polarization of the afterglow of GRB\,990712 which seems to vary
with time, while the polarization angle does not change with time (Rol
et al. 2000b). These observations cannot be easily reconciled with
afterglow polarization theories. The photometric measurements show a
common temporal power-law decay of the transient source, overtaken by
the bright host galaxy at late times; no evidence is found for a
supernova of type SN1998bw (Sahu et al. 2000; see also Hjorth et
al. 2000).

The organization of this paper is as follows: in \S 2 we present the
observations and data reduction methods.  In \S 3 we display and
discuss the spectra of GRB\,990510, followed by GRB\,990712 in \S
4. We study the absorption-line intensity evolution in time for both
bursts in \S 5 and describe our conclusions in \S 6.

\section{Observations}

After the optical identification of both GRB\,990510 (Vreeswijk et
al. 1999) and GRB\,990712 (Bakos et al. 1999), we triggered our VLT
target-of-opportunity observation program and obtained several
low-resolution spectra at various epochs with the FOcal Reducer and
low-dispersion Spectrograph (FORS), mounted at the Cassegrain focus of
the ESO VLT-UT1 {\it Antu} telescope.  The date of observation, grism
used, wavelength range, resolving power and the signal-to-noise at
6500\AA\ are listed in Table \ref{tab:log}. A slit width of 1\arcsec\
was used for all spectra. Grism G150I approximately covers the
wavelength range 3700-9000 \AA. However, redward of 6500 \AA\ the
second order starts to contaminate the first order (cf. FORS User
Manual 1.3). To obtain a clean spectrum longward of 6500 \AA\ for
GRB\,990510, we also took spectra with an order separation filter
(OG590), using the same grism. However, due to the low sensitivity of
the CCD shortward of 3700 \AA, the impact of the overlap is negligible
shortward of 7700 \AA. Therefore, we have summed the blue and red
spectra over the region 6200-7700 \AA\ and combined this part with the
single blue and red spectra into a continuous spectrum over the entire
wavelength range. For the grisms G300V and G300I, we have simply
connected the blue and red parts into one spectrum. For GRB\,990712,
we used grism G150I without order separation filter, and thus these
spectra are usable over the wavelength range 3700-7700 \AA.

The raw spectra were bias-subtracted, and flat-fielded with a
normalized combined set of lamp flat-fields. Subsequently, cosmic rays
were removed interactively along the afterglow spectrum and each
sequence of images was summed into combined images.  The spectra were
optimally extracted from the combined images, and wavelength
calibrated using a standard Helium-Neon-Argon lamp. The r.m.s. error
in the wavelength calibration is approximately 0.25 \AA. Since no
useful spectra of standard stars were taken neither during the first
two nights for GRB\,990510 nor for GRB\,990712, we have
flux-calibrated these spectra using the BVRI light curve data of Rol
et al. (2000a) for GRB\,990510 and the VRI data of Sahu et al. (2000)
for GRB\,990712. The spectra of the other nights were flux-calibrated
with a spectrophotometric standard star, which resulted in flux levels
that are consistent with the photometry at the same epochs.

To flux-calibrate the spectrum of GRB\,990510, we fitted the light
curves with a smoothly connected broken-power-law model (Harrison et
al. 1999, Stanek et al. 1999, Beuermann et al. 1999), while for
GRB\,990712 we used a simple power-law model with a host galaxy
contribution. We determined the magnitudes at the times when the
spectra were taken (see Table \ref{tab:log}), using the fits to the
light curves. These values were then corrected for the estimated
Galactic foreground extinction: E(B-V) = 0.2 and E(B-V) = 0.03
(Schlegel, Finkbeiner \& Davis 1998) for the May and July burst,
respectively.  The E(B-V) value is translated into an extinction at a
given wavelength using the standard Galactic extinction curve of
Cardelli, Clayton \& Mathis (1989). The magnitudes were transformed to
fluxes (Fukugita, Shimasaku \& Ichikawa 1995) and were fitted with a
power-law spectrum F $\propto \nu^{\beta}$. For GRB\,990510 we find
$\beta$ = --0.6 $\pm$ 0.1 (first night), and $\beta$ = --0.7 $\pm$ 0.1
(second night). For the three spectra of GRB\,990712 we obtain
$\beta$(1) = --1.1 $\pm$ 0.2, $\beta$(2) = --0.9 $\pm$ 0.2 and
$\beta$(3) --0.9 $\pm$ 0.2. We have not taken into account the
extinction intrinsic to the host galaxy; see \S 4. These slopes are
quite usual for GRB afterglows.  We also fitted the global profile
(excluding the absorption and emission lines) of all the
wavelength-calibrated spectra with a 4th-order Chebychev
polynomial. To obtain the flux-calibrated spectra, we multiplied the
wavelength-calibrated spectra by the ratio between the power-law fit
based on the photometry and the global spectral profile fit. We
estimate the error in the flux calibration for all nights to be about
15\%.

The equivalent width (EW) of the spectral lines was measured using the
{\it splot} routine in IRAF; we used both a Gaussian fit, and also
simply summed the difference between the pixel value and the continuum
for each pixel over the line.  Both methods gave similar results. The
error is mostly dominated by the uncertainty in the continuum level,
which was estimated by placing the continuum at a high and low level
(roughly corresponding to the $\pm$ r.m.s. value of the continuum in
the vicinity of the line). We take the mean value of these two
estimates as the EW, and half their difference as the error. We also
calculate the Poisson error from the object and sky spectrum, which we
quadratically sum with the measurement error to obtain the total
error. We then corrected the EW for the host galaxy contribution in
case of absorption features (i.e. divided the measured EW by the
afterglow fraction of the total light) and for the afterglow
contribution in case of an emission line (i.e. divided by the
galaxy-light fraction of the total). The host galaxy fraction of the
total light for GRB\,990510 is negligible (see \S 3), while for
GRB\,990712 we obtain 0.23 $\pm$ 0.03, 0.30 $\pm$ 0.04 and 0.47 $\pm$
0.06 for epoch one, two and three, respectively, from the V and R band
light curve fits of Sahu et al. (2000). The error was estimated from
the different values for the host galaxy magnitude as given by Sahu et
al. (2000) and Hjorth et al. (2000). These errors are propagated along
with the errors in the EW measurements. Finally the EW is converted to
the rest frame by dividing by (1+z). The emission-line fluxes are
independent of the afterglow contribution to the total light.

\section{The absorption-line spectrum of GRB\,990510}

We have identified several absorption lines in the first epoch
spectrum of GRB\,990510: Mg{\sc ii} \l 2800, Mg{\sc i} \l 2853, Fe{\sc
ii} \l\l 2344,2383,2600, and possibly Al{\sc iii} \l 1683 and Cr{\sc
ii}/Zn{\sc ii} \l 2062. The observed wavelengths, identifications,
corresponding redshifts, and equivalent widths (in the absorber's rest
frame) are listed in Table \ref{tab:510lines}. The lines and the
telluric absorption features are also indicated in Figure 1.

\begin{deluxetable}{cccl}
\footnotesize
\tablecaption{Absorption lines in GRB\,990510 \label{tab:510lines}}
\tablecomments{$W_{\lambda}$ is given in the absorber's rest frame, i.e. the
observed value is divided by (1+z). The error in $W_{\lambda}$ is
dominated by the uncertainty in the placing of the continuum. The
relative error expected from Poisson statistics and read noise is
given in parentheses. The weighted mean redshift is
calculated, using the separate measurements of all the lines, even
though the identification of Al{\sc iii} and Cr{\sc ii}/Zn{\sc ii} is
uncertain (see text).}
\tablehead{
\colhead{$\lambda_{\rm obs}$ (\AA)} & 
\colhead{ID} &
\colhead{z$_{\rm abs}$} &
\colhead{{$W_{\lambda,\rm rest}$} (\AA)}}
\startdata
4883   $\pm$ 6   & Al{\sc iii}(1862.79)?	 		&  1.6213 $\pm$ 0.0032 & 0.8 $\pm$ 0.6 (0.010) \\
5396   $\pm$ 5	 & Cr{\sc ii}(2062.23)/Zn{\sc ii}(2062.66)? 	&  1.6166 $\pm$ 0.0024 & 0.8 $\pm$ 0.3 (0.010) \\
6142   $\pm$ 3   & Fe{\sc ii}(2344.21)		 		&  1.6201 $\pm$ 0.0013 & 0.6 $\pm$ 0.2 (0.011) \\
6241   $\pm$ 5	 & Fe{\sc ii}(2382.76)		 		&  1.6192 $\pm$ 0.0021 & 0.5 $\pm$ 0.3 (0.012) \\
6806   $\pm$ 3   & Fe{\sc ii}(2600.17) 		 		&  1.6175 $\pm$ 0.0011 & 0.9 $\pm$ 0.2 (0.011) \\
7335   $\pm$ 2   & Mg{\sc ii}(2796.35/2803.53) 	 		&  1.6197 $\pm$ 0.0034 & 2.6 $\pm$ 0.4 (0.017) \\
7472   $\pm$ 6	 & Mg{\sc i}(2852.96)		 		&  1.6190 $\pm$ 0.0021 & 0.6 $\pm$ 0.2 (0.013) \\
\tableline
		 &	weighted mean:	&  1.6187 $\pm$ 0.0015 & \\
\enddata
\end{deluxetable}

\begin{figure*}[tbp]
%\null\hspace{-0.5cm}
\centerline{\psfig{file=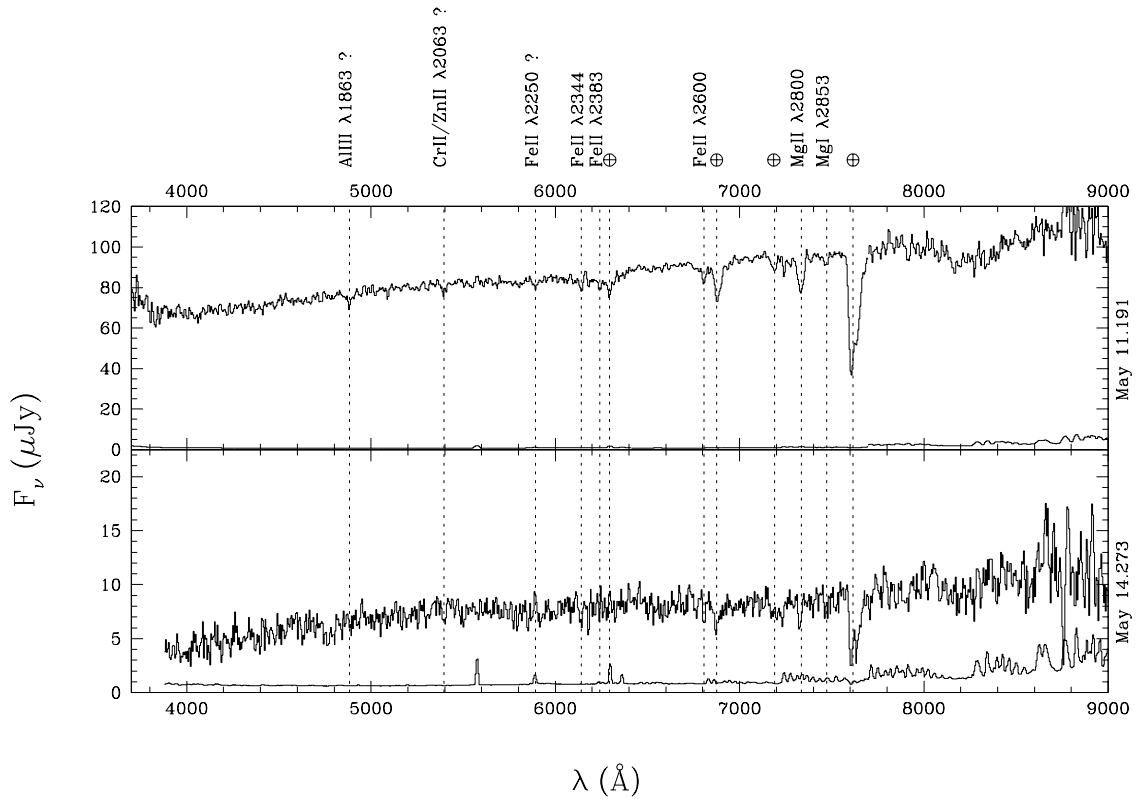,width=6in,angle=0}}
\figcaption{VLT/FORS1 spectra of the afterglow of GRB\,990510; 0.8 and
3.9 days after the burst. The error spectra, calculated from the
object and sky spectra using Poisson statistics, are also shown.
Several absorption features (including telluric lines marked with
$\oplus$) are indicated with the broken lines.}
\label{fig:510spectra}
\end{figure*}

These line identifications can be verified by taking into account the
oscillator strength, ionization potential and relative cosmic
abundance.  The observed EWs of Fe{\sc ii} at 2344 \AA, 2383 \AA, and
2600 \AA\, are in reasonable agreement with their relative oscillator
strengths (Morton 1991). Other Fe{\sc ii} lines, such as Fe{\sc ii} \l
1608 and Fe{\sc ii} \l 2587 were not observed, but these have smaller
oscillator strengths, consistent with their non-detection. However,
the oscillator strength of the undetected Al{\sc iii} \l 1855 is twice
that of the detected Al{\sc iii} \l 1863 line, and so should have been
detected at an observed wavelength of 4858 \AA. This makes the
identification of the line at 4883 \AA\ with Al{\sc iii}
questionable. A similar argument can be made for the reality of the
Cr{\sc ii}/Zn{\sc ii} \l 2062 line.  Both lines are expected to be
accompanied by a stronger partner (Cr{\sc ii} \l 2056 and Zn{\sc ii}
\l 2026), which is not detected.  The presence of Fe{\sc ii} suggests
that the medium probed with our line of sight is most likely one of
low-ionization, and therefore Al{\sc ii} \l 1670 is expected to be
present as well given the similarity between the Fe and Al ionization
potentials, but again nothing is detected. However, we clearly detect
Mg{\sc ii}, which is a blend of Mg{\sc ii} \l 2796 and Mg{\sc ii} \l
2804, and observe Mg{\sc i} \l 2853 in absorption as well.

The weighted mean redshift of the identified lines is $z = 1.6187 \pm$
0.0015. On the basis of this redshift we can possibly identify Fe{\sc
ii}(2250) at 5893 \AA, although we do not expect to detect this line
due to its low oscillator strength. Since no clear emission lines are
detected, this redshift is certainly a lower limit to the redshift of
the GRB afterglow. However, the strength of Mg{\sc i} \l 2853 suggests
we are probing a dense, low-ionization medium, very likely that of the
host galaxy. Assuming isotropic emission, $z$ = 1.619, H$_0$ = 70 km
s$^{-1}$ Mpc$^{-1}$, $\Omega_0$=0.3, and $\Lambda$=0, we find a
luminosity distance of 3.5$\times$10$^{28}$cm, corresponding to an
($>$ 20 keV) energy output of 1.5 $\times 10^{53}$ erg for
GRB\,990510, based on the BATSE fluence of 2.56$\times$10$^{-5}$ erg
cm$^{-2}$ (Kippen et al. 1999; see also Briggs et al. 2000).

Unfortunately, the spectra taken on May 12 and May 16 (not shown in
Fig. 1) are of inferior quality, and cannot be used
for absorption-line measurements. In the May 14 spectrum, however, we
detect Mg{\sc ii} again. The observed wavelength of $\lambda_{\rm
obs}$ = 7327 $\pm$ 5 \AA\ is consistent with the redshift determined
in the first epoch spectrum. Its equivalent width is $W_{\rm rest}$ =
2.3 $\pm$ 0.6 \AA, which is (within the errors) identical to the
measurement of May 11.2 1999 UT (0.8 days after the burst; $W_{\rm
rest}$ = 2.6 $\pm$ 0.4 \AA). For the other absorption lines detected
earlier it is not possible to obtain an accurate measurement of their
EW.

We will now derive a lower limit to the H{\sc i} column in the
direction of GRB\,990510 from the Fe{\sc ii} lines at 2344 \AA, 2600
\AA, and 2383 \AA\ in the first epoch spectrum. The relation between
the EW of a line, $W_{\lambda}$, its column density, $N_{j}$ (number
of atoms in the corresponding ionization state), and the oscillator
strength, {\it f}, of the transition (j) is (Spitzer 1978):

log $\frac{W_{\lambda}}{\lambda} = {\rm log}(N_{j}\times\lambda\times f) -
4.053$ 

\noindent

Here the unit of $W_{\lambda}$ is \AA, the column density $N_{j}$ is
in $\rm{cm^{-2}}$ and the wavelength $\lambda$ is in cm. We do not use
the strongest absorption line, Mg{\sc ii} \l 2800, since this line is
easily saturated in typical galaxy spectra, placing it on the flat
part of the curve of growth ($W_{\lambda}$ versus $N_{j} \times
\lambda \times f$). Even if the error in the EW determination were
very small, this would still lead to a very uncertain value for the
column density. The Mg{\sc i} line is not used either, due to the
large uncertainty in the ratio of Mg{\sc i} to Mg{\sc ii}, translating
in a similar uncertainty in the column density. Fe{\sc ii}, however,
should be the dominant Fe ion in a dense neutral environment. The
following holds if the Fe lines are not saturated. Using the
oscillator strength values from Morton (1991) for these lines ({\it
f}(Fe{\sc ii} \l 2344)=0.110, {\it f}(Fe{\sc ii} \l 2383)=0.301 and
{\it f}(Fe{\sc ii} \l 2600)=0.224), and the values for $W_{\lambda}$
obtained from the spectrum of the first epoch (see Table
\ref{tab:510lines}), we obtain log $N$(Fe{\sc ii} \l 2344) = 14.0
$\pm$ 0.1 $\rm{cm^{-2}}$, log $N$(Fe{\sc ii} \l 2383) = 13.5 $\pm$ 0.2
$\rm{cm^{-2}}$ and log $N$(Fe{\sc ii} \l 2600) = 13.8 $\pm$ 0.1
$\rm{cm^{-2}}$. We adopt the value log $N$(Fe) = 13.8 $\pm$ 0.2
$\rm{cm^{-2}}$. Even though the EWs of the three different Fe lines
result in similar values for the column, the lines are probably
saturated, i.e. our estimate for the Fe column density should be
considered as a lower limit.

In converting the number of Fe atoms to a hydrogen column density, we
have to take into account that the metallicity in high-redshift
galaxies is likely to be lower than the Galactic value, and that a
large fraction of the Fe atoms can be hidden in dust (Whittet 1992).
A study of the metallicity in damped Ly$\alpha$ systems (which have
$N$(H{\sc i}) $\gsim$ 10$^{20} \rm{cm^{-2}}$) from $z$ = 0.7 to 3.4
(Pettini et al. 1997; we adopt [Zn/H]=--0.8), allows us to estimate
the Fe abundance at the redshift of GRB\,990510 with respect to the
solar abundance (Grevesse \& Sauval 1999; log(Fe/H)$_{\odot}$ =
--4.5), obtaining log($N$(Fe)/$N$(H))= --5.3. The Galactic Fe
depletion factor in a cool disk environment is --2.2 dex (Sembach \&
Savage 1996, Savage \& Sembach 1996), but here we adopt the typical
depletion measured in damped Ly$\alpha$ systems, which is around --0.6
dex. Taking all these points into account gives the following rough
lower limit to the hydrogen column density: log $N$(H{\sc i}) $\geq$
19.7 $\rm{cm^{-2}}$. Another and more robust way of estimating the
H{\sc i} column density which is independent of dust corrections, is
to use the Fe{\sc ii} measured in DLAs (the gas phase only) around the
redshift of the GRB. Using 10 systems with redshifts ranging from 1.2
to 2.0, we find [Fe/H] = --1.5 $\pm$ 0.5. Assuming that this is the
most likely [Fe/H] abundance for the host galaxy of GRB\,990510, we
obtain log $N$(H{\sc i}) $\geq$ 13.8 + 1.5 + 4.5 $\rm{cm^{-2}}$ = 19.8
$\rm{cm^{-2}}$, very close to our first estimate. A low column density
is consistent with that found by Briggs et al. (2000) from fitting a
combined set of BATSE and BeppoSAX X-ray and $\gamma$-ray data.

HST imaging, performed in April 2000, appears to detect a very faint
(V $\sim$ 28) galaxy, located only 0.08\arcsec\ East of the position
of the early afterglow (Fruchter et al. 2000, Bloom et al.
2000). This galaxy is probably responsible for the detected absorption
lines in the spectra. The type and strength of the absorption lines,
which are indicative of a low-ionization, high-density medium,
strongly suggest that these originate in the host galaxy of
GRB\,990510. Even though the galaxy is extremely faint, the GRB
afterglow allows us to probe its interstellar medium and -- in
principle -- to measure its metallicity.  For the latter we need a
measure of the column density that does not depend on the strength of
the metal absorption lines (e.g. from a Balmer feature).

\section{The emission- and absorption-line spectrum of GRB\,990712}

Fig. 2 shows the spectra of GRB\,990712, taken at
0.5, 0.7 and 1.5 days after the burst (see Table \ref{tab:log}).  The
obvious emission lines are easily identified as [O{\sc ii}] \l 3727,
[Ne{\sc iii}] \l 3869, H$\gamma$, H$\delta$, and [O{\sc iii}] \l\l
4959,5007. We also detect two absorption lines, which can be
identified as Mg{\sc ii} \l 2796,2804 and Mg{\sc i} \l 2853 at the
same redshift as the emission lines, and are therefore intrinsic to
the host galaxy. Using all these features in all three spectra, we
obtain a weighted average redshift of $z$ = 0.4331 $\pm$ 0.0004. In
Table \ref{tab:712lines} we list the observed wavelength of the
features, their identification and redshift, and the rest frame EW and
flux (corrected for the Galactic foreground extinction), for the
spectra taken at different epochs. The last two columns contain the EW
and flux values averaged over all the spectra.

\begin{deluxetable}{llcrrrrrrrr}
\footnotesize
%\tabletypesize{\scriptsize}
%\rotate 
%\null\hspace{-2cm}
\tablewidth{19cm}
\tablecaption{Absorption and emission lines in GRB\,990712 \label{tab:712lines}}
%\tablecomments{}
\tablehead{
\colhead{} & \colhead{} & \colhead{} & \colhead{July 13.18} & \colhead{} & \colhead{July 13.42} & \colhead{} & \colhead{July 14.18} & \colhead{} & \colhead{weighted mean} & \\
\colhead{$\lambda_{\rm obs}$ (\AA)} & \colhead{ID} & \colhead{z} & \colhead{{$W_{\lambda,\rm rest}$} (\AA)} & \colhead{{Line flux$^{\rm a,b}$}} & \colhead{{$W_{\lambda,\rm rest}$} (\AA)} & \colhead{{Line flux}} & \colhead{{$W_{\lambda,\rm rest}$} (\AA)} & \colhead{{Line flux}} & \colhead{$W_{\lambda,\rm rest}$} & \colhead{{Line flux}}
}
\startdata
4015 $\pm$ 6 & Mg{\sc ii} \l\l 2796,2804  & 0.4340 $\pm$ 0.0021 &   8.3 $\pm$  1.4 &                 &   9.7 $\pm$  1.9 &                 & 13.7 $\pm$  4.5 & & 9.1 $\pm$ 1.3 & \\ 	
4096 $\pm$ 6 & Mg{\sc i} \l 2853  	   	& 0.4357 $\pm$ 0.0021 &   2.2  $\pm$  0.6 &                 &  $<$ 2 (2$\sigma$)&                &  3.9 $\pm$  1.4 & & 2.5 $\pm$ 0.7 & \\ 	
5342 $\pm$ 3 & [O{\sc ii}] \l 3727        	& 0.4332 $\pm$ 0.0008 & --47.1  $\pm$  7.4 & 3.27 $\pm$ 0.20 &  --44.7 $\pm$  7.9 & 3.33 $\pm$ 0.22 &  --45.8 $\pm$  8.6 & 3.37 $\pm$ 0.33 & --45.9 $\pm$ 1.2 & 3.32 $\pm$ 0.14 \\
5545 $\pm$ 5 & [Ne{\sc iii}] \l 3869         	& 0.4332 $\pm$ 0.0013 &  --8.4  $\pm$  1.5 & 0.60 $\pm$ 0.08 &  --10.9 $\pm$  2.3 & 0.79 $\pm$ 0.11 &   --7.5 $\pm$  1.9 & 0.55 $\pm$ 0.11 & --8.6 $\pm$ 1.2 & 0.63 $\pm$ 0.10 \\
6220 $\pm$ 3 & H$\gamma$ \l 4340            	& 0.4332 $\pm$ 0.0007 &  --8.1  $\pm$  3.3 & 0.50 $\pm$ 0.18 &   --6.0 $\pm$  2.5 & 0.39 $\pm$ 0.14 &   --4.5 $\pm$  1.8 & 0.29 $\pm$ 0.10 & --5.5 $\pm$ 1.3 & 0.35 $\pm$ 0.11 \\
6966 $\pm$ 3 & H$\beta$ \l 4861          	& 0.4330 $\pm$ 0.0006 & --21.1  $\pm$  3.9 & 1.17 $\pm$ 0.12 &  --28.6 $\pm$  5.3 & 1.59 $\pm$ 0.17 &  --22.4 $\pm$  7.7 & 1.23 $\pm$ 0.31 & --23.5 $\pm$ 3.5 & 1.33 $\pm$ 0.20 \\
7106 $\pm$ 3 & [O{\sc iii}] \l 4959         	& 0.4330 $\pm$ 0.0006 & --43.7  $\pm$  7.0 & 2.36 $\pm$ 0.18 &  --40.7 $\pm$  6.2 & 2.28 $\pm$ 0.16 &  --42.8 $\pm$  8.8 & 2.38 $\pm$ 0.30 & --42.2 $\pm$ 1.5 & 2.32 $\pm$ 0.12 \\
7175 $\pm$ 2 & [O{\sc iii}] \l 5007         	& 0.4330 $\pm$ 0.0004 & --113.5 $\pm$ 17.1 & 6.04 $\pm$ 0.31 & --110.7 $\pm$ 15.1 & 6.15 $\pm$ 0.25 & --108.0 $\pm$ 18.6 & 6.18 $\pm$ 0.53 & --110.9 $\pm$ 10.2 & 6.14 $\pm$ 0.19 \\
\tableline
             & weighted mean:	  		&  0.4331 $\pm$ 0.0004 & &&&&&\\
\enddata
\tablenotetext{a}{All flux units: 10$^{-16}$ erg s$^{-1}$ cm$^{-2}$}
\tablenotetext{b}{All line fluxes have been corrected for the Galactic
foreground extinction.}
\end{deluxetable}

\normalsize
\begin{figure*}[tbp]
%\null\vspace{-1cm}
%\null\hspace{-0.5cm}
\centerline{\psfig{file=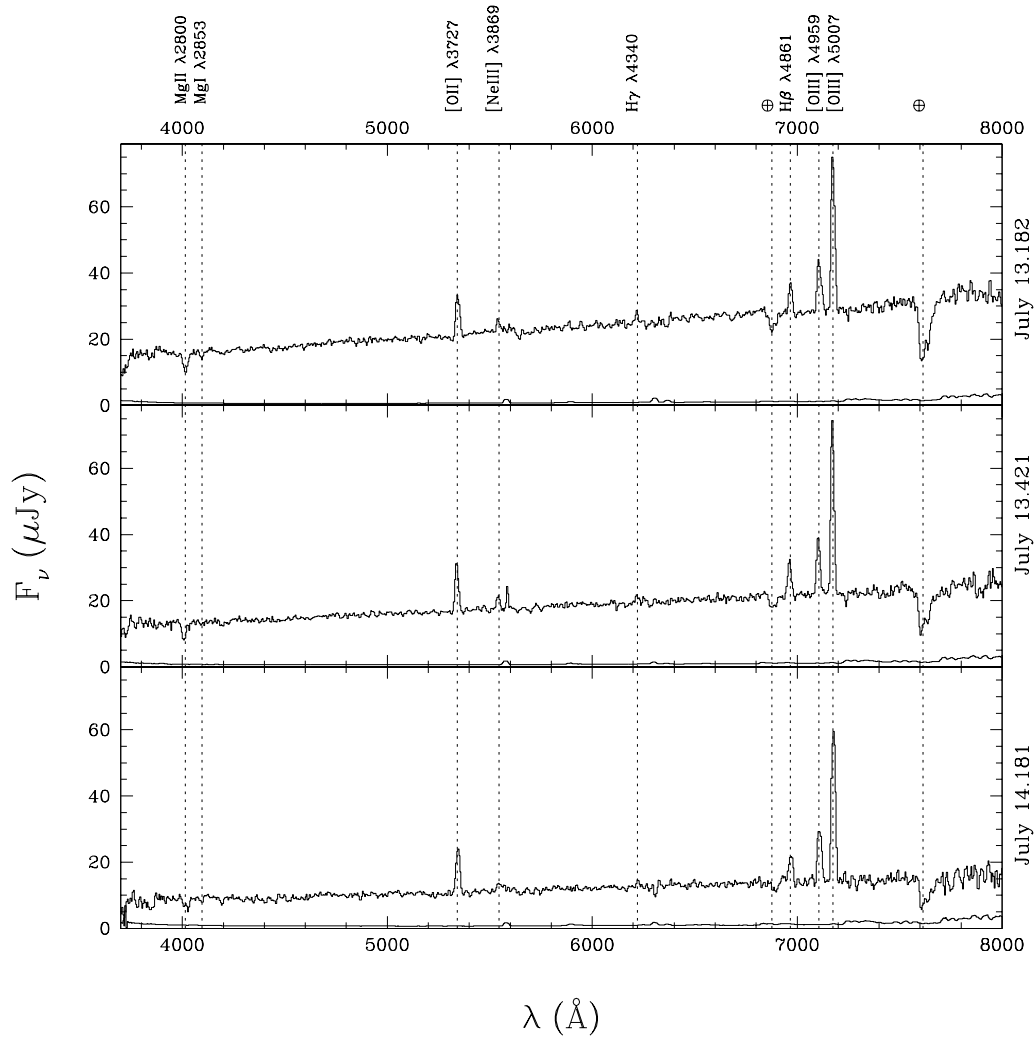,width=5.5in,angle=0}}
\caption{VLT/FORS1 spectra of the afterglow of GRB\,990712, from 0.5
to 1.5 days after the burst. The error spectra, calculated from the
object and sky spectra using Poisson statistics, are also shown.  The
absorption features (including telluric lines marked with $\oplus$)
and emission features are indicated with the broken lines. The
observation dates (also shown in the figure), grisms, resolving power
and signal-to-noise ratios are listed in Table \ref{tab:log}.}
\label{fig:712spectra}
\end{figure*}

It is important to investigate whether GRB host galaxies are indeed
star forming galaxies, i.e. H{\sc ii} galaxies, where massive O and B
stars ionize the interstellar medium, giving rise to prominent
emission lines. These lines are also observed in galaxies that host an
active galactic nucleus (AGN, e.g. Seyfert 2), where their strength
does not only depend on the star formation, but on the nuclear
activity as well. Most popular GRB progenitor models require a close
connection with massive-star formation, but so far, this has not been
confirmed for any of the host galaxies.

Hjorth et al. (2000) suggested that the host galaxy of GRB\,990712 may
be a Seyfert 2 galaxy on the basis of the [O{\sc iii}] \l
5007/H$\beta$ ratio being greater than three (see Shuder \& Osterbrock
1981), and its location in the log([O{\sc iii}] \l 5007/H$\beta$)
vs. log([O{\sc ii}]/[O{\sc iii}] \l 5007) diagram (see Fig. 2 of
Baldwin, Phillips \& Terlevich 1981). We also measure a ratio of
[O{\sc iii}] \l 5007/H$\beta$ that is greater than three: 4.6 $\pm$
1.0. However, from the more recent work of Rola, Terlevich \&
Terlevich (1997), who employ the Canada-France Redshift Survey (CFRS)
sample of emission-line galaxies at redshifts 0 $< z \leq$ 0.3, it is
clear that a value of 4.6 is actually very typical for H{\sc ii}
galaxies (see their Fig. 1). Combined with our value for log([O{\sc
ii}]/H$\beta$) of 0.4 $\pm$ 0.1, the host of GRB\,990712 is located
clearly within the H{\sc ii} galaxy regime. We note that this diagram
is corrected for extinction intrinsic to the distant galaxies, while
the values in our Table \ref{tab:712lines} are not. However, the
[O{\sc iii}] \l 5007/H$\beta$ ratio is only slightly affected by
reddening. Rola et al. (1997) also build non-extinction corrected
diagrams to distinguish between H{\sc ii} galaxies and other
emission-line galaxies; in all but one of these, where its location is
on the border of the H{\sc ii} galaxy and Seyfert 2 regimes, the host
galaxy of GRB\,990712 is classified as an H{\sc ii} galaxy (e.g. Table
\ref{tab:712lines} gives C$_{3727}$--C$_{4861}$ = --0.28 $\pm$ 0.03,
whereas Fig. 4 of Rola et al. 1997 shows all definite Seyfert 2s have
C$_{3727}$--C$_{4861} >$ 0.4), i.e. the emission lines are produced by
H{\sc ii} regions that are being ionized by O and B stars. We conclude
GRB\,990712 is most likely an H{\sc ii} galaxy, and not a Seyfert 2.

We have estimated the host galaxy extinction, by comparing the
observed ratio of H$\gamma$/H$\beta$ (0.26 $\pm$ 0.09) with the
expected ratio for case B recombination (0.469 $\pm$ 0.009; Osterbrock
1989). Using the Galactic extinction curve of Cardelli et al. (1989),
we obtain A$_V$ = \gpm{3.4}{2.4}{1.7}. The relative large error is due
to the marginal detection of H$\gamma$.

We can now estimate the star formation rate (SFR) in the host galaxy
of GRB\,990712 in three different ways: through the [O{\sc ii}] and
H$\beta$ line luminosities (Kennicutt 1998), and through the continuum
flux at 2800 \AA\ (Madau et al. 1998). For all these estimates a
Salpeter initial mass function (IMF) has been assumed. Combining
equation (3) of Kennicutt (1998), a luminosity distance of 6.8
$\times$ 10$^{27}$ cm (taking H$_0$ = 70 km s$^{-1}$ Mpc$^{-1}$,
$\Omega_0$=0.3, and $\Lambda$=0), and the [O{\sc ii}] \l 3727 line
flux from Table \ref{tab:712lines} (the fluxes in this table have
already been corrected for Galactic extinction) we obtain SFR([O{\sc
ii}]) = 2.7 $\pm$ 0.8 M$_{\odot}$ yr$^{-1}$, consistent with the value
found by Hjorth et al. (2000). Taking into account the measured
extinction (at H$\alpha$, due to the fact that the [O{\sc ii}] SFR
method is calibrated through H$\alpha$), using the Galactic extinction
curve (Cardelli et al. 1989), we find SFR([O{\sc ii}]) =
\gpm{35}{178}{25} M$_{\odot}$ yr$^{-1}$.  Using equation (2) of
Kennicutt (1998), the case B (large optical depth) line ratio {\it
j}$_{H\alpha}$/{\it j}$_{H\beta}$ of 2.85 (Osterbrock 1989) and the
H$\beta$ flux of Table \ref{tab:712lines}, we find SFR(H$\beta$) 1.7
$\pm$ 0.6 M$_{\odot}$ yr$^{-1}$. Corrected for reddening, this becomes
SFR(H$\beta$) = \gpm{64}{770}{54} M$_{\odot}$ yr$^{-1}$.  Finally,
equation (2) of Madau et al. (1998), combined with the 2800 \AA\ flux
(which is located at 4013 \AA\ at a redshift of 0.4331), gives
SFR(2800 \AA) = 2.8 $\pm$ 0.9 M$_{\odot}$ yr$^{-1}$, which becomes
$\sim$ 400 M$_{\odot}$ yr$^{-1}$, using A$_V$ = 3.4. The [O{\sc ii}]
method, although it is indirectly calibrated through H$\alpha$ and
sensitive to the abundance and ionization state of the gas, is
probably the least uncertain of the three. The H$\beta$ estimate is
uncertain due to the unknown host galaxy stellar absorption underneath
the emission.  The SFR based on the ultra-violet continuum is very
uncertain due to the fact that our flux calibration is an
extrapolation below the V band, and, more importantly, the extinction
correction at ultra-violet wavelengths is very uncertain. Normally,
the 2800 \AA\ SFR is found to be lower by a factor of 2--3 (at
redshifts up to $z$ = 2.8) as compared to the H$\alpha$ luminosity SFR
method (Glazebrook et al. 1999, Yan et al. 1999).

For the GRB host galaxies for which a SFR has been determined so far,
even for the [O{\sc ii}] method alone, the values range from 0.5 $\pm$
0.15 M$_{\odot}$ yr$^{-1}$ (GRB\,970228; Djorgovski et al.  1999a) to
20 $\pm$ 9 M$_{\odot}$ yr$^{-1}$ (GRB\,980703; Djorgovski et
al. 1998); host galaxy extinction is not included, so these values
should be considered as lower limits. Normalized by the host B band
luminosity, this range is narrowed down to spread a factor of three.
It has been noted (e.g. Djorgovski et al. 1998), that the range of
SFRs for GRB host galaxies does not seem to be extra-ordinarily high,
as compared to field galaxies at similar redshifts. E.g. Glazebrook et
al. (1999) find a range of 20-60 M$_{\odot}$ yr$^{-1}$ (using
H$\alpha$), for 13 field galaxies at $z$ = 1, drawn from the
CFRS. However, the galaxies at high redshifts for which these rates
have been measured tend to be much brighter than the typical GRB host
galaxy, and are therefore expected to have much higher SFRs. The
[O{\sc ii}] equivalent width, which effectively is the star formation
rate normalized by the blue band luminosity of the host galaxy, allows
a more useful comparison. The mean $W_{\rm rest}$([O{\sc ii}]) of the
Glazebrook et al. sample is 33 with a standard deviation of 15 \AA,
while we measure $W_{\rm rest}$([O{\sc ii}]) = 46 $\pm$ 1 \AA\ for the
host of GRB\,990712. This comparison also suggests that GRB\,990712
occurred in a galaxy that is vigorously forming stars.

Using the oxygen and H$\beta$ emission lines, we can estimate the
global oxygen abundance of the host of GRB\,990712 (see Kobulnicky,
Kennicutt \& Pizagno 1999, and references therein). From the values in
Table \ref{tab:712lines} we obtain $R_{23} \equiv (I_{3727} + I_{4959}
+ I_{5007})/H\beta = 8.9 \pm 1.5$, corresponding to log(O/H) = --3.7
$\pm$ 0.4 (we note that log(O/H)$_{\odot}$ = --3.1; Cox et al. 1999).
This estimate is based on fluxes not corrected for reddening, but such
a correction would not change the abundance estimate
substantially. For comparison: the oxygen abundances for a sample of
22 relatively nearby spiral galaxies range from log(O/H) = --2.7 to
--3.5 (Kobulnicky, Kennicutt \& Pizagno 1999).

\section{The time dependence of the {\rm Mg{\sc ii}} feature}

It is now well established that GRBs are the most energetic events in
the universe, with peak isotropic luminosities up to 10$^{53}$ erg
s$^{-1}$ (Kulkarni et al. 1998). Also when the radiation from GRBs is
beamed, the impact of the shock on the circumburst material is the
same as in an isotropic explosion, and could be observable as
time-resolved evolution of the Fe K$\alpha$ line and edge in the X-ray
regime (Weth et al. 1999, and references therein) and of ultra-violet
(UV) absorption lines, redshifted to the optical domain (Perna \& Loeb
1998). By monitoring the line evolution it is possible to obtain
information on the density structure surrounding the explosion, and --
in case the density can be measured by an independent method -- the
redshift to the burst may be obtained (Perna \& Loeb 1998).

Our spectral observations of the afterglows of GRB\,990510 and
GRB\,990712 extend over several nights, and both contain absorption
lines, so that we can look for possible temporal evolution of these
features. They are expected to decrease in time, if a considerable
fraction of the atoms responsible for the absorption are in the
vicinity of the site of the burst, and are ionized by the explosion
(Perna \& Loeb 1998). Alternatively, the burst may release atoms that
are locked in the dust, which could result in a corresponding increase
of the EWs. The clearest absorption feature in both bursts is Mg{\sc
ii}.  In the May 11 and May 14 spectra of GRB\,990510 (0.8 and 3.9
days after the burst, respectively), we measure $W_{\rm rest}$ = 2.6
$\pm$ 0.4 \AA\ (May 11) and $W_{\rm rest}$ = 2.3 $\pm$ 0.6 \AA\ (May
14). For GRB\,990712 we obtain $W_{\rm rest}$ = 8.3 $\pm$ 1.4 \AA,
$W_{\rm rest}$ = 9.7 $\pm$ 1.9 \AA\ and $W_{\rm rest}$ = 13.7 $\pm$
4.5 \AA\ for 0.5, 0.7 and 1.5 days after the burst, respectively. For
both bursts, the values are constant within the errors.  This is not
very surprising; the Mg{\sc ii} feature is most likely saturated in
the spectra of both bursts. This means that over a wide range of
column densities, the EW is not expected to change by a detectable
amount. On the other hand, a significant change in the EW would have
indicated a large change in the column density.  The Mg{\sc i} feature
in GRB\,990712, which is possibly saturated as well, is also constant
within the errors.

\section{Conclusions}

GRB afterglows allow us to probe galaxies that would otherwise be
extremely difficult or impossible to study spectroscopically. We have
determined a lower limit to the redshift of GRB\,990510 through
identification of several spectral absorption lines. The strength of
the Mg{\sc i} line is indicative of a cool, dense environment, which
leads to the conclusion that the measured $z$ = 1.6187 $\pm$ 0.0015
most likely reflects absorption in the host galaxy ISM.

Using both the absorption and emission lines in the spectrum of
GRB\,990712, we determine its redshift at $z$ = 0.4331 $\pm$ 0.0004.
The emission-line ratios indicate that the host of GRB\,990712 is an
H{\sc ii} galaxy, with an [O{\sc ii}] star formation rate (reddening
corrected) of \gpm{35}{178}{25} M$_{\odot}$ yr$^{-1}$. The large
[O{\sc ii}] equivalent width, compared to that of field galaxies at $z
\leq$ 1, also suggests that the host is vigorously forming stars.

In order to put meaningful constraints on the circumsource medium,
high resolution, high signal-to-noise spectra at several epochs after
the burst are needed to resolve the velocity structures of the
non-saturated lines, and allow determination of the circumsource
density distribution and its evolution. This may become possible with
the launch of HETE-II, since this satellite will allow follow-up
observations of optical transients at early times, i.e. when they are
bright (R $\sim$ 16). Determination of the density profile could
provide a major advance in solving the progenitor problem.

\acknowledgements PMV and ER are supported by the NWO Spinoza
grant. CK acknowledges support from NASA grant NAG 5-2560.  TJG
acknowledges support from the Sherman Fairchild Foundation.  LK is
supported by a fellowship of the Royal Academy of Sciences of the
Netherlands. We especially want to thank the ESO/VLT staff on Cerro
Paranal who performed most of the TOO observations and assisted us
during the nights of FORS Consortium guaranteed time.

\end{document}